# Quantum Channel AlGaN/GaN/AlGaN High Electron Mobility Transistor

G. Simin[1] and M. Shur[2,3][a]

[1] *Department of Electrical Engineering, University of South Carolina, Columbia SC, 29208, USA*

[2] *Rensselaer Polytechnic Institute, Troy, NY, 12180, USA, Email: shurm@rpi.edu  hhttps://orcid.org/0000-0003-0976-6232*

[3]*Electronics of the Future, Inc. Vienna, Va 22181, USA*

Scaling down the GaN channel in a double heterostructure AlGaN/GaN/AlGaN High Electron Mobility Transistor (HEMT) to the thicknesses on the order of or even smaller than the Bohr radius confines electrons in the quantum well even at low sheet carrier densities. In contrast to the conventional designs, this Quantum Channel (QC) confinement is controlled by epilayer design and the polarization field and not by the electron sheet density. As a result, the breakdown field at low sheet carrier densities increases by approximately 36% or even more because the quantization leads to an effective increase in the energy gap. In addition, better confinement increases the electron mobility at low sheet carrier densities by approximately 50%. Another advantage is the possibility of increasing the aluminum molar fraction in the barrier layer because a very thin layer prevents material relaxation and the development of dislocation arrays. This makes the QC especially suitable for high-voltage, high-frequency, high-temperature, and radiation-hard applications.

Ever since their emergence in the early 1990s [1, 2] GaN High Electron Mobility Transistors have demonstrated superior performance for high voltage [3, 4], high frequency [5,6], high temperature [7,8] operation, and excellent radiation hard potential [9,10]. The prime reasons for their superiority for operation in extreme environments are a wide energy gap leading to a high breakdown voltage and polarization doping [11, 12] supporting large sheet electron densities in the HEMT channels. A high polar optical phonon energy in GaN (91.2 meV [13] compared to 35 meV for GaAs [14]) and screening of impurity scattering by a large electron density ensure a high field effect mobility, especially in the on-state and lead to a low on-resistance for long [15-17] and short channel devices.[18]

In this letter, we show that these advantages could be dramatically enhanced using the Quantum Channel HEMT (QCHEMT) design (see Fig. 1). QCHEMT incorporates a thin channel comparable in thickness to the Bohr radius ($a_B$=2.14 nm for GaN) sandwiched between the top and bottom wide bandgap barriers. Fig. 1 compares QCHEMT and conventional GaN HEMT designs. QC-HEMT is a radical transformation of Double heterostructure III-Nitride HEMTs (DHEMTs) that demonstrated superior performance. [[19-22]] In the DHEMT design, the role of the bottom back barrier was to improve electron confinement and reduce electron trapping in the depth of the GaN layer. In contrast to conventional GaN HEMT designs, the shape of the potential confining two-dimensional electron gas (2DEG) in QC-HEMT is determined by both the polarization field and bandgap discontinuities at the top and bottom barrier interface. In QCHEMT quantization effects are strongly pronounced with the ground energy state substantially above the bottom of the conduction band. Fig. 2 compares the energy band diagrams of a conventional AlGaN/GaN HEMT, double-heterostructure HEMT (DHFET), and the novel QC HEMT with a channel thickness of 2 nm.

---

[a] G. Simin and M. Shur contributed equally to this work

QC HEMT design leads to two important results: first, the two-dimensional electron gas in the device channel is confined at gate voltages close to or even slightly below the threshold and second, the ground state in the quantum well remains well above the bottom of the conduction band in GaN. The ground state energy can be found as [[23]]

$$E_o = \left(\frac{\hbar^2}{2m}\right)^{1/3} \left(\frac{3\pi}{2} qF_{eff}\right)^{2/3} (n_q + 3/4)^{2/3} \qquad (1)$$

where $n_q$ is the quantum number ($n_q = 0$ for the ground state), and $F_{eff}$ is the effective electric field in the channel. Our simulations show that, just like for Si MOSFETs[24], the effective electric field in the conventional HEMT is approximately

$$F_{eff} \approx \frac{F_i}{2} = \frac{qn_s}{2\epsilon\varepsilon_o}, \qquad (2)$$

Here $F_i$ is the electric field at the barrier-channel interface, $n_s$ is the sheet electron density in the channel, $\varepsilon_o$ is the dielectric permittivity of vacuum, and $\varepsilon$ =8.9 is the GaN dielectric constant. Factor 2 in the denominator accounts for the band bending. [23]

For the QC HEMT $F_{eff} \approx F_s + \frac{F_i}{2}$, where $F_s$ is the polarization field that depends on the molar fractions of the cladding layers and the QC thickness. We used a self-consistent solution of the Schrodinger-Poisson equation to estimate $F_s = 10^8$ V/m. Fig. 3 compares the dependencies of the ground state energy $E_0$ above the bottom of the conduction band on the sheet carrier concentration $n_s$ for the regular HEMT and for the QC HEMT with a 2 nm thick channel generated using Eqs. (1-3). We have also found the ground energy states for QC HEMT using a 1D Poisson simulator by G. Shnider [25] and obtained very close results.

As seen, the position of the lowest quantum state $E_0$ in the quantum channel device remains practically constant even at $n_s$ at the gate voltage close to the threshold. This is equivalent to the effective increase in the energy gap by $E_0$. Note that for the breakdown field consideration, the device parameters at the gate voltage close to the threshold are particularly important. The analysis presented in reference [26] shows that the breakdown field, $F_{BR}$, is approximately proportional to $E_G^{2.5}$, consistent with the experimental data presented in [27]. For QC HEMT, the effective energy gap $E_{Geff} = E_G + E_0$. For the QC HEMT of Fig 2(b), $E_G$ =3.39 eV, $E_0$=0.45 eV at gate voltage close to the threshold, and the resulting increase in an $F_{BR}$ is around 36%.

Strong electron confinement in QC HEMT may lead to another important effect resulting in further $F_{BR}$ increase. As shown in [28] at high electron energies, real space transfer of hot electrons from the quantum well into the barrier should occur. Fig. 4 shows the simulated electron wave function for the QC HEMT of Fig. 2(b). As seen, strong electron confinement leads to a significant fraction of electrons penetrating the top barrier. Therefore, it is reasonable to expect that at a high electric field, a large fraction of the channel would experience quantum transfer to the top barrier. Since the top barrier is made of material with a larger bandgap (in our example, it is AlGaN with 65% Al), a further significant increase in the effective breakdown field $F_{BR}$ is expected. This effect requires further theoretical and experimental studies.

Strong electric confinement in QC HEMT leads to a smaller 2DEG effective $\Delta d$ thickness than it is for a conventional HEMT. The $\Delta d$ could be estimated as the ratio of the ground state energy over the electric field at the heterointerface:

$$\Delta d = \frac{E_0}{qF_i} \qquad (3)$$

Hence, bulk (volume) electron density for the same $n_s$ value is higher in QC HEMT as compared to conventional HEMT. Fig. 5 shows the volume electron density as a function of $n_s$ in conventional and QC HEMT calculated using Eqs. (1 – 3). As seen, for conventional HEMT, volume electron density rapidly decreases as $n_s$ decreases, i.e., as the gate bias approaches the threshold because the effective width of the 2DEG $\Delta d$ in conventional HEMT rapidly increases as the gate bias approaches the threshold; 2DEG confinement nearly disappears. In QC HEMT, the volume electron density is a much slower function of $n_s$. A relative increase in volume density leads to better screening and less impurity scattering, hence to higher mobility. Fig. 6a illustrates this expected improvement extracted from the measured data [29].

Experimental confirmation of mobility increase due to better confinement in double heterostructure (DH) HEMTs has been obtained earlier in [21]. This data is shown in Fig. 6(b). The field effect mobility in III-Nitride HEMTs depends on many factors such as interface roughness, defect concentration, strain, alloy scattering, and dislocation density. However, strong scattering screening due to higher concentration should result in higher mobility in any HEMT. QC HEMT with a 2 nm thick channel exhibits nearly 10 times higher electron confinement than in the DH HEMT reported in [21].

In conclusion, the QC AlGaN/GaN/AlGaN HEMT design should considerably improve the breakdown voltage, field-effect mobility, and, therefore, transconductance at low electron sheet densities and support a large maximum electron sheet density and a larger maximum current. A thinner quantum well made of lower bandgap material should also lead to better radiation hardness of QC HEMT. The proposed QC HEMT concept can be applied to other material compositions, e.g., to AlGaN channel HEMT which will lead to even higher breakdown voltages without sacrificing electron mobility.

Data Availability Statement

The data that support the findings of this study are available from the corresponding author upon reasonable request.

Acknowledgement

We gratefully acknowledge support from the Office of Naval Research (contract # No N00014-23-1-2289).

**Figures**

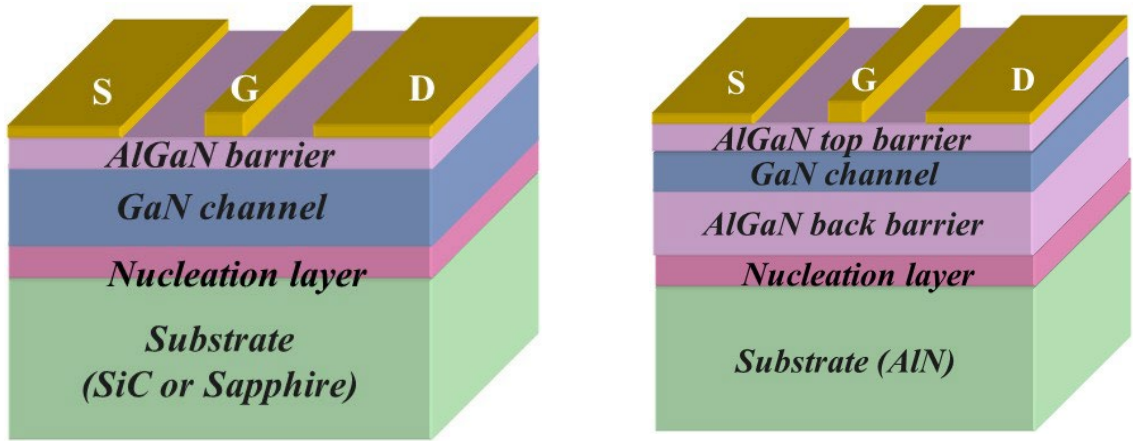

(a)            (b)

Fig. 1. Conventional (a) and QC double heterostructure (b) HEMT designs

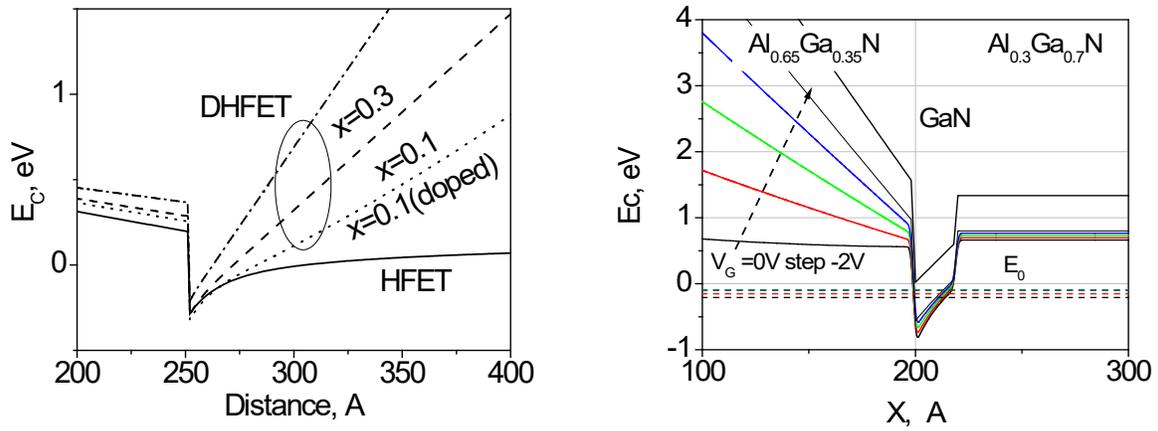

Fig. 2. Energy band diagrams of (a) conventional AlGaN/GaN HEMT, double-heterostructure HEMT (DHFET) with 20 nm channel [21] and (b) novel QC HEMT with channel thickness 2 nm. The dashed line in (b) shows ground state energy $E_0$ at different gate voltages.

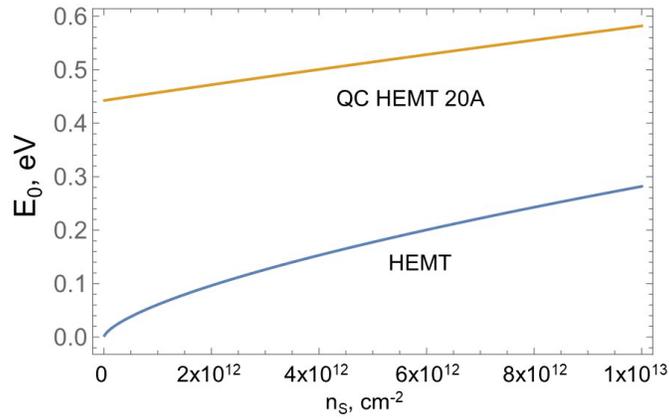

Fig. 3. Dependencies of the ground state energy $E_0$ above the bottom of the conduction band on the sheet carrier concentration $n_s$ for the conventional HEMT and the QC HEMT with a 2 nm thick channel.

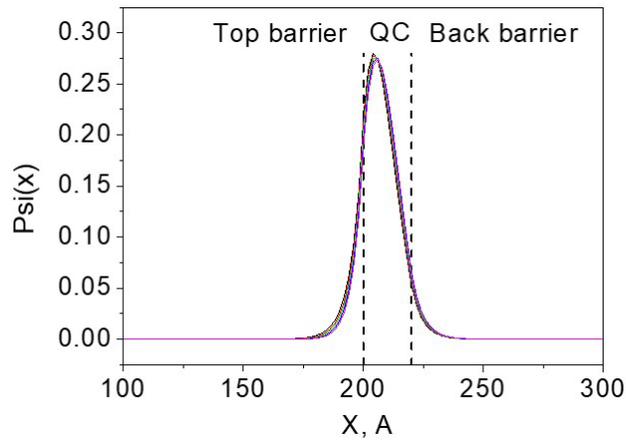

Fig. 4. Electron wave function for the QC HEMT with Fig. 3(b) with 20A thick channel. Different curves correspond to different gate voltages. As seen, the wave function profile is practically independent of $V_G$.

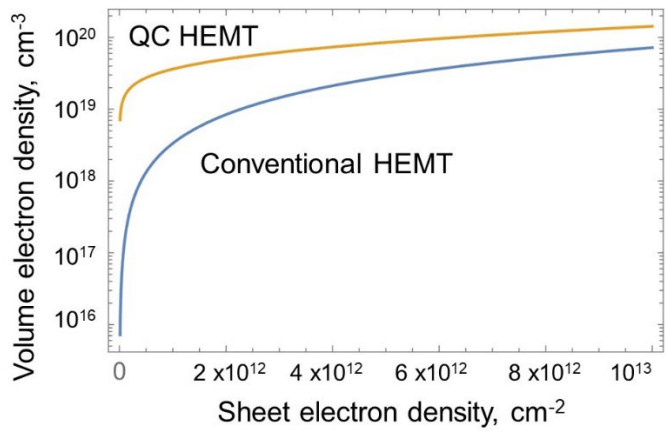

Fig. 5. 2DEG volume electron density as a function of $n_s$ in conventional and QC HEMTs.

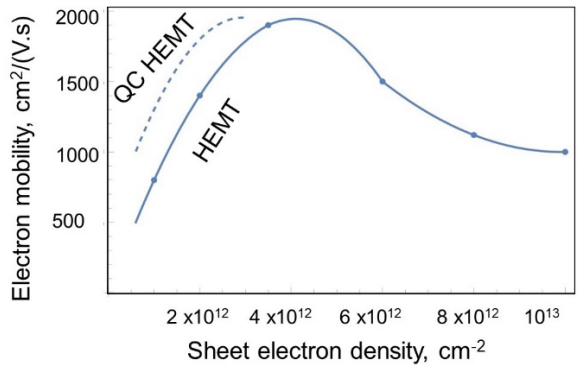 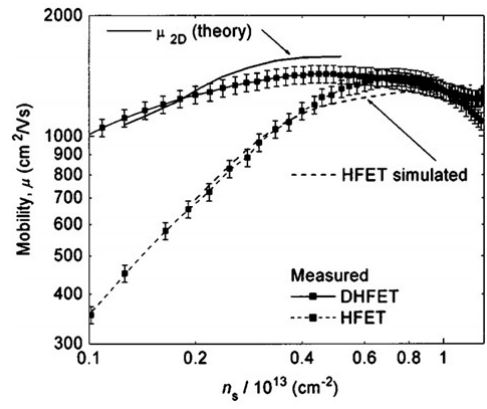

(a) (b)

Fig. 6. (a) Expected mobility increase in QC HEMT; (b) experimentally observed mobility increase in double-heterostructure (DH) HEMT.